\documentclass[%
reprint,
amsmath,amssymb,
aps,
]{revtex4-1}

\usepackage{graphicx}
\usepackage{appendix}
\usepackage{varioref}
\usepackage{cleveref}
\usepackage{subfig}
\usepackage{dcolumn}
\usepackage{bm}
\usepackage[mathlines]{lineno}

\usepackage{xcolor}
\begin{document}
	
	
	\title{Direct and Indirect Methods of Vortex Identification in Continuum limit}
	
	\author{Sedigheh\ Deldar}%
	\email{{\color{blue}{sdeldar@ut.ac.ir}}}
	\author{Zahra\ Asmaee}
	\email{{\color{blue}{zahra.asmaee@ut.ac.ir}}}
	\affiliation{%
		Department of  Physics, University of Tehran,\\
		P. O. Box 14395/547, Tehran 1439955961, Iran
	}%
	\begin{abstract}
	Inspired by direct and indirect maximal center gauge methods which identify vortices in lattice calculations, and by using the connection formalism, 
	we show that under some appropriate gauge transformations, vortices and chains appear in the continuum limit of QCD vacuum.

		\vspace*{0.2cm}
		\begin{description}
			\item[PACS numbers]
			12.38.Aw, 12.38.Lg, 12.40.-y.
		\end{description}
	\end{abstract}

	\pacs{Valid PACS appear here}
	\maketitle
	
	
\section{\label{sec:level1}Introduction}
Confinement mechanism is one of the most controversial unsolved issues in particle physics in the low-energy regime or large distances. 
In order to study the confinement potential between a pair of quark and anti-quark, the Quenched approximation is used where the dynamical quarks are removed for the infrared regime.
In fact, we can obtain some collective modes from the gluons \cite{Ichie} which are associated with some topological degrees of freedom of the QCD vacuum. Magnetic monopoles and center 
vortices are among the main candidates for describing the confinement problem.

In the absence of matter fields, the center vortex model has been suggested as a possible mechanism of confinement to extract some degrees of freedom of pure Yang-Mills theory. The idea is that the QCD vacuum is filled with closed magnetic vortices, and it is assumed that the vortices are condensed in it. 
Vortices are defined by the center of the SU($N$) gauge group, and there exist $(N-1)$ distinct vortices, which are called non-Abelian Z$_{N}$ vortices. They produce full string tensions as the Yang-Mills vacuum does. 

To study the confinement problem by center vortices, one first has to discuss the existence of vortices in the continuum limit.
The most common methods of identifying vortices in the lattice simulations are direct maximal center gauge (DMCG) \cite{DFG98} and indirect maximal center gauge (IMCG) \cite{DFG97}.
Using these two methods and by the help of the connection 
formalism \cite{Ichie}, we discuss about the appearance of vortices in the continuum limit of QCD.

\section{\label{sec:level2}The Direct Method of Introducing Vortices in SU($2$) Gauge Group}

Motivated by DMCG  method in lattice QCD, we study vortices in the continuum limit. 
In this method, the formation of center vortices in the QCD vacuum relies upon two steps: center gauge transformation and center projection. 
\\
In the continuum limit, the gluon field is transformed as
\begin{equation}
	\vec{A}^G_\mu.\vec{T}=G(x)\left( A^c_{\mu}T^c\right)  G^{\dagger}(x)-\dfrac{i}{g} G(x)\partial_\mu G^{\dagger}(x),
	\label{1}
\end{equation}
where $\vec{A}^G_\mu.\vec{T}\in SU(N)$, and $T^c$ are generators of the SU($N$) group and $c$ is the color index.
After the center gauge transformation, thin vortices appear as topological defects.
Therefore, Eq.\eqref{1} can be used for studying the vortices if $G(x)\equiv N(x)$ is defined as a center gauge transformation.
On the other side, we recall that in lattice QCD calculations when
the Wilson loop links to the vortex it receives a phase difference equal to $e^{i2\pi n/N}$ associated with the non-trivial center element contribution Z$(k)$.
Therefore, under a center gauge transformation $N(x)$, a Wilson line should be transformed as \cite{ER2017},
\begin{equation}
	\begin{split}
		W^N(C^{\prime})&=N(x-\epsilon)W(C^{\prime})N^{\dagger}(x+\epsilon)\\
		&=N(x-\epsilon)N^{\dagger}(x+\epsilon)+\mathcal{O}(\epsilon)
		\equiv Z(k)+\mathcal{O}(\epsilon),
		\label{2}
\end{split}\end{equation}
where $W(C^{\prime})=1+\mathcal{O}(\epsilon)$. $C^{\prime}$ indicates an open circle from $x-\epsilon$ to $x+\epsilon$, where $x$ indicates the location of the intersection
of $C^{\prime}$ and the hypersurface $\Sigma$. $\epsilon$ is an infinitesimal quantity so that in the limit where $\epsilon \rightarrow 0$, $C^{\prime}=C$.
Eq.\eqref{2} must be used to find an appropriate center gauge transformation.

We recall that an ideal vortex is defined on $(D-1)$-dimensional hypersurface $\Sigma$ vortex, while the thin vortex is defined on $(D-2)$-dimensional boundary $S=\partial\Sigma$. Piercing the hypersurface vortex by the Wilson loop, results in a discontinuity Z$(k)$. 
The relation between an ideal vortex and a thin vortex is the following \cite{ER2017},
\begin{equation}
	\text{ideal\ vortex}=-\dfrac{i}{g}N(x)\partial_\mu N^{\dagger}(x)-\text{thin\ vortex}.
	\label{3}
\end{equation}
Replacing Eq.\eqref{3} in Eq.\eqref{1} for $G(x)\equiv N(x)$, one gets 
\begin{equation}
	\vec{A}^N_\mu.\vec{T}=N(x)\left( A^c_{\mu}T^c\right)  N^{\dagger}(x)+\text{ideal\ vortex}+\text{thin\ vortex}.
\end{equation}
On the other hand, as observed in lattice calculation, it is the thin vortex that links to the Wilson loop.
One can define a gauge field in the coset space by removing the ideal vortex \cite{ER2017}, so that $\vec{A}^{ N}_\mu.\vec{T}\rightarrow \vec{A}^{\prime N}_\mu.\vec{T}$. Finally, we get
\begin{equation}
	\vec{A}^{\prime N}_\mu.\vec{T}=N(x)\left( A^c_{\mu}T^c\right)  N^{\dagger}(x)+\text{thin\ vortex}.
	\label{4}
\end{equation}
For $x\notin$ hypersurface, we only see the boundary of the vortex, called the thin vortex field. Thus, the contribution of the ideal vortex would be zero:
\begin{equation}
	\text{thin\ vortex}=-\dfrac{i}{g} N(x)\partial_\mu N^{\dagger}(x),\quad x\notin \text{hypersurface}.
	\label{5}
\end{equation}
For SU($2$) case, a gauge transformation is written in terms of three Euler angles $\alpha$,$\beta$,$\gamma$,
\begin{equation}
	\begin{split}
		&G(x)=e^{i\gamma(x)T^3}e^{i\beta(x)T^2}e^{i\alpha(x)T^3},
		\\
		&\alpha(x)\in\left[ 0,2\pi\right),\ \beta(x)\in \left[ 0,\pi\right],\ \gamma(x)\in\left[ 0,2\pi\right),\ T^c=\dfrac{\sigma^c}{2},
		\label{6}
	\end{split}
\end{equation}
where $\sigma^c \left( c=1,2,3\right) $ are the Pauli matrices.
The center gauge transformation $G(x)\equiv N(x)\in SU(2)$ is continuous everywhere except at the hypersurface of the vortex.
Therefore, the Euler angles are selected in a way that the constraint of Eq.\eqref{4} is satisfied. 
There are different choices for the angles. One can choose $\alpha=\gamma=\dfrac{\varphi}{2}$ and $\beta=0$ and thus,
\begin{equation}
	N=e^{i\varphi T^3}, \quad \varphi\in\left[ 0,2\pi\right).
	\label{7}
\end{equation}
It can be shown that for $\varphi=0$, $N(\varphi=\epsilon)N^{\dagger}(\varphi=2\pi-\epsilon)=-\textbf{1}_{2\times 2}\in Z(2)$ when $\epsilon\rightarrow 0$. Thus the contribution of an ideal vortex is observed at $\varphi=0$. On the other hand, outside the hypersurface, the contribution of the thin vortex can be identified as a pure gauge shown in  Eq.\eqref{5},
\begin{equation}
	\text{thin\ vortex}\equiv  \vec{V}_\mu.\vec{T}=-\dfrac{i}{g}N\partial_\mu N^{\dagger}=-\dfrac{1}{g}\partial_\mu\varphi T^3.
	\label{8}
\end{equation}
In cylindrical coordinates, the thin vortex is observed explicitly at $\rho=0$ \cite{ER2017} in the third direction of the color space, $\vec{V}_{\varphi}.\vec{T}=-\dfrac{g^{-1}}{\rho}T^3$. 
The magnetic vortex flux is $\Phi^{\text{flux}}=\int dx^{\mu}\left( \vec{V}_{\mu}.\vec{T}\right)=-\dfrac{2\pi}{g}T^3.$ 
Under the center gauge transformation Eq.\eqref{7}, the gluon field Eq.\eqref{4} is written in terms of the local color frame $\hat{n}_c$, so that 
$NT^cN^{\dagger}\equiv \hat{n}_c.\vec{T}$ and the vector potential is transformed as,
\begin{equation}
	\vec{A}^{\prime N}_\mu=A^1_\mu\hat{n}_1+A^2_\mu\hat{n}_2+\left( A^3_\mu-\dfrac{1}{g}\partial_\mu\varphi\right)\hat{k},
	\label{10}
\end{equation}
where $\hat{n}_3=\hat{k}$. Therefore some topological defects appear as a result of the singular gauge transformation. To observe these defects explicitly, 
we rewrite the field strength in terms of the covariant-derivative operator $\hat{D}_\mu$ and the ordinary derivative operator $\hat{\partial}_\mu$ $(\text{See \cite{Ichie}})$,
\begin{equation}
	F_{\mu\nu}=\dfrac{1}{ig}\left[ \hat{D}_\mu,\hat{D}_\nu\right]-\dfrac{1}{ig}\left[ \hat{\partial}_\mu, \hat{\partial}_\nu\right],
	\label{11}
\end{equation}
where $F_{\mu\nu}$ is the SU($N$) non-Abelian field strength tensor, and Eq.\eqref{11} is applied when the singularity exists. 
It should be noted that for the regular systems, the second term in the right hand of Eq.\eqref{11} is zero.

In general, if $G(x)\in SU(N)$ represents a gauge transformation, the field strength tensor is transformed as $F^G_{\mu\nu}=G(x)F_{\mu\nu}G^{\dagger}(x)$. If we use a singular system defined by $F_{\mu\nu}$ in Eq.\eqref{11} we have,
\begin{equation}
	F^G_{\mu\nu}= \left( \partial_\mu A^{ G}_\nu-\partial_\nu A^{G}_\mu\right)  +ig\left[A^{ G}_\mu,A^{G}_\nu \right]+\dfrac{i}{g}G\left[ {\partial}_\mu, {\partial}_\nu\right]G^{\dagger}.
	\label{12}
\end{equation}
This is called the connection formalism technique and as a result of using this technique, the theory will remain gauge invariant after the singular gauge 
transformation.

For SU($2$) case, by the help of Eq.\eqref{10}, we rewrite the first term of Eq.\eqref{12} for the gauge transformation $G(x)\equiv N(x)$. The first term is linear in terms of $\vec{A}^{\prime N}_{\mu}$,
\begin{equation}\begin{split}
		\vec{F}^{\text{linear}}_{\mu\nu}&=\sum_{c=1}^{3}\left( \partial_\mu A^c_\nu-\partial_\nu A^c_\mu\right)\hat{n}_c+\left( \partial_\mu V_{\nu}- \partial_\nu V_{\mu}\right)\hat{k}
		\\
		&-g\left(A^1_\nu V_{\mu}-A^1_\mu V_{\nu} \right) \hat{n}_2+g\left(A^2_\nu V_{\mu}-A^2_\mu V_{\nu} \right) \hat{n}_1.
		\label{13}
\end{split} \end{equation}
The first  term of Eq.\eqref{13} is regular  and the second term represents the field strength of a thin vortex field carrying a magnetic flux equal to $\Phi^{\text{flux}}=-\dfrac{2\pi}{g}T^3$. The third and the fourth terms indicate some kind of interactions between thin vortex and the off-diagonal gluon fields. 
Using Eq.\eqref{10}, the second term of Eq.\eqref{12}, $\vec{F}^{\text{bilinear}}_{\mu\nu}\equiv \dfrac{i}{g}\left[ \vec{A}^{\prime N}_\mu,\vec{A}^{\prime N}_\nu\right] $  can be written in terms of the local frame $\hat{n}_c$,
\begin{equation}\begin{split}
		\vec{F}^{\text{bilinear}}_{\mu\nu}&=-g\sum_{a,b,c=1}^{3}\varepsilon^{abc}\left( A^a_\mu A^b_\nu-A^b_\mu A^a_\nu\right)\hat{n}_c
		\\
		&+g\left(A^1_\nu V_{\mu}-A^1_\mu V_{\nu} \right) \hat{n}_2-g\left(A^2_\nu V_{\mu}-A^2_\mu V_{\nu} \right) \hat{n}_1.
		\label{14}
\end{split} \end{equation}
The first term of Eq.\eqref{14} represents interactions between gluon fields, and is regular. The second and the third terms indicate interactions between the thin 
vortex and the off-diagonal gluon fields but with an opposite sign compared with their counterparts in Eq.\eqref{13}.
The third term of Eq.\eqref{12}, $\vec{F}^{\text{singular}}_{\mu\nu}\equiv \dfrac{i}{g}N\left[ {\partial}_\mu, {\partial}_\nu\right]N^{\dagger}=-\left( \partial_\mu V_{\nu}- \partial_\nu V_{\mu}\right)\hat{k}$ 
indicates  the field strength of an anti-thin vortex carrying magnetic flux of $\Phi^{\text{flux}}=+\dfrac{2\pi}{g}T^3$.
The anti-thin vortex field  strength tensor contribution represented by $\vec{F}^{\text{singular}}_{\mu\nu}$ is canceled by the thin vortex field strength tensor contribution brought in the 
second term of $\vec{F}^{\text{linear}}_{\mu\nu}$; and finally one is left with a full QCD field strength tensor.
In fact, with the above parametrization, we have shown that the vacuum is filled with vortices and anti-vortices.
Therefore, it is clear that if one wants to have only the contribution of vortices, one has to discard the term $\vec{F}^{\text{singular}}_{\mu\nu}$. Thus, 
\textquotedblleft the center projected\textquotedblright\ field strength tensor is defined as the following,
\begin{equation}
	\vec{F}^{\text{CP}}_{\mu\nu}\equiv \vec{F}^{\text{linear}}_{\mu\nu}+\vec{F}^{\text{bilinear}}_{\mu\nu}.
	\label{15}
\end{equation}
To summarize, we have directly shown that under a center gauge 
transformation followed by a \textquotedblleft center projection\textquotedblright, a gauge field configuration is obtained which contains a thin vortex.
\section{\label{sec:level3}The Indirect Method of  Introducing Vortices in SU($2$) Group}

Another method of introducing vortices in lattice QCD is the IMCG method and we are interested in investigating this method in the continuum limit. We show that by 
indirect method, we would have chains of vortices and monopoles in comparison with section \eqref{sec:level2}, where we have only a single vortex.
In this method, in addition to the center gauge transformation and center projection, an initial step called Abelian gauge transformation and then Abelian projection is used \cite{DFG97}.
Choosing  $\alpha(x)=\varphi$, $\beta(x)=\theta$ and $\gamma(x)=\pm\varphi$ for the gauge rotation matrix of Eq.\eqref{6}, one gets an appropriate 
$G\equiv M\in SU(2)$ which leads to an Abelian gauge transformation. In this paper, we choose $\gamma(x)=-\varphi$,
\begin{equation}
	M(\theta,\varphi)=e^{-i\varphi T^3}e^{i\theta T^2}e^{i\varphi T^3}.
	\label{17}
\end{equation} 
We define $M(\theta,\varphi)$ as an Abelian gauge transformation, then the transformation of the gluon field is given by Eq.\eqref{1},
\begin{equation}
	\vec{A}^M_\mu.\vec{T}=M\left( \sum^3_{c=1}A^c_{\mu}T^c\right)  M^{\dagger}-\dfrac{i}{g} M\partial_\mu M^{\dagger}.
	\label{18}
\end{equation}
The first term on the right hand side of Eq.\eqref{18} is regular under Abelian gauge transformation $M$, but the second term is singular and in the spherical coordinates is obtained by,
\begin{equation}\begin{split}
		\textbf{A}&^{\text{singular}}(\theta,\varphi)=\dfrac{g^{-1}}{r}\left(cos\varphi \textbf{e}_{\varphi}+sin\varphi \textbf{e}_{\theta} \right) T^1
		\\&+\dfrac{g^{-1}}{r}\left(sin\varphi \textbf{e}_{\varphi}-cos\varphi \textbf{e}_{\theta} \right) T^2+\dfrac{g^{-1}}{r}\dfrac{1-cos\theta}{sin\theta}\textbf{e}_{\varphi}T^3,
		\label{19}
\end{split}\end{equation}
where $\textbf{A}^{\text{singular}}(\theta,\varphi)=\textbf{A}^{c\ \text{singular}}(\theta,\varphi)T^c$. It is observed from Eq.\eqref{19} that there exists a magnetic monopole as a point defect at the origin, $r=0$ along with a Dirac string at $\theta=\pi$.
The magnetic flux $\Phi^{\text{flux}}(\theta)$ of the singular term is $\Phi^{\text{flux}}(\theta)=\dfrac{2\pi}{g} \left(1-cos\theta\right)T^3$. At $\theta=\pi$, 
the magnetic flux of a Dirac string that enters a monopole located at the origin $r=0$, is equal to $\dfrac{4\pi}{g}T^3$.

It is clear that under the Abelian gauge transformation, the field strength tensor has a form similar to Eq.\eqref{12}, but with the $G\equiv M$ 
which indicates the Abelian gauge transformation.
Since the two color directions $T^1$ and $T^2$ have no contribution in the magnetic flux, we suppress these non-diagonal components of the gauge fields in the infrared regime and use only the diagonal sector.
It can be easily confirmed that the first term of Eq.\eqref{12} for the Abelian sector $\left( F^{\text{linear}}_{\mu\nu}\right)^3 \equiv \partial_\mu\left(A^M_\nu\right)^3-\partial_\nu\left( A^M_\mu\right)^3$ includes a magnetic monopole sitting at the origin along with a Dirac string in $\theta=\pi$.
The second term of Eq.\eqref{12} for the Abelian sector, 
$\left( F^{\text{bilinear}}_{\mu\nu}\right)^3 \equiv ig\left\lbrace \left( A^M_\mu\right)^1\left( A^M_\nu\right)^2-\left(A^M_\mu\right)^2\left( A^M_\nu\right)^1 \right\rbrace $
contains an anti-monopole at the origin, and the third term of Eq.\eqref{12} for the Abelian sector, 
$\left( F^{\text{singular}}_{\mu\nu}\right)^3\equiv\dfrac{i}{g}M(\theta,\varphi)\left[ {\partial}_\mu, {\partial}_\nu\right]M^{\dagger}(\theta,\varphi)$, includes an anti-Dirac string at $\theta=\pi$ with a magnetic flux equal to $-\dfrac{4\pi}{g}T^3$.

The sum of the two terms $F^{\text{linear}}_{\mu\nu}+F^{\text{singular}}_{\mu\nu}$ represents a gauge configuration that only contains a monopole at $r=0$. However it is exactly canceled by the anti-monopole arisen from $F^{\text{bilinear}}_{\mu\nu}$, such that a field strength tensor which gives a full QCD is obtained. Thus, one can claim that the vacuum is filled with monopoles and anti-monopoles.
Therefore, in order to have only the contribution of the monopole, we define the projected gauge fields as,
\begin{equation}
	\vec{A}^M_\mu.\vec{T}=\left( A^M_\mu\right)^aT^a \rightarrow \mathcal{A}_\mu\equiv \left( A^M_\mu \right) ^3 T^3,
	\label{20}
\end{equation}
where,
\begin{equation}
	\left( A^M_\mu \right) ^3 =A^{\text{regular}}_{\mu}+\dfrac{1}{g}\left( 1-cos\theta\right)\partial_\mu\varphi.
	\label{21}
\end{equation}
As a result, $F^{\text{bilinear}}_{\mu\nu}$  which represents the anti-monopole contribution is equal to zero and the remaining 
part $F^{\text{linear}}_{\mu\nu}+F^{\text{singular}}_{\mu\nu}$ describes an Abelian projection QCD which contains a monopole at $r=0$, and is called the monopole vacuum.
Next, we have to do a center gauge transformation on the monopole vacuum. Similar to the arguments of section \eqref{sec:level2}, we obtain
\begin{equation}
	\vec{A}^{\prime NM}_\mu.\vec{T}=N(x) \mathcal{A}_\mu N^{\dagger}(x)+\text{thin\ vortex}.
	\label{22}
\end{equation}
Using the center gauge transformation defined in Eq.\eqref{7} and Eq.\eqref{20}, one obtains
\begin{equation}
	\vec{A}^{\prime NM}_\mu.\vec{T}=\left[ A^{\prime\ \text{regular}}_\mu+\dfrac{1}{g}\left( 1-cos\theta\right)  \partial_\mu\varphi-\dfrac{1}{g}\partial_\mu\varphi\right] T^3.
	\label{23}
\end{equation}
$E_\mu\equiv-\dfrac{1}{g}cos\theta \partial_\mu\varphi T^3$ indicates a defect representing a monopole located at the origin $r=0$ along with the two 
line vortices at $\theta=0,\pi$ in spherical coordinates.
In fact, the magnetic potential of the chain defined by $E_\mu \equiv-\dfrac{1}{g}cos\theta \partial_\mu\varphi T^3$, can 
be interpreted as the sum of two terms: a magnetic potential of a monopole along with a Dirac string defined by 
$B_\mu\equiv \dfrac{1}{g}\left( 1-cos\theta\right) \partial_\mu\varphi T^3$ plus a magnetic potential of a 
vortex defined by $V_\mu\equiv -\dfrac{1}{g}\partial_\mu\varphi T^3$.
Therefore, the chain flux is obtained as the sum of the vortex flux and the
magnetic monopole flux plus the Dirac string and it is equal to $-\dfrac{2\pi}{g}cos\theta T^3$.
Now, we present some discussions about the chain characteristic. For $\theta=0$, we only have the contribution of a magnetic line vortex flux equal to $-\dfrac{2\pi}{g}T^3$ located in the positive direction of the $z$-axis which enters to the magnetic monopole placed at the origin, $r=0$. 
At $\theta=\pi$, there exists a Dirac string flux equal to $+\dfrac{4\pi}{g}T^3$ located in the negative direction of the $z$-axis and enters the magnetic monopole. There is also a magnetic line vortex whose flux is equal to $-\dfrac{2\pi}{g}T^3$ at $\theta=\pi$. It is located in the negative direction of the $z$-axis and exits from the magnetic monopole placed at $r=0$.
In fact, the sum of the two fluxes $\Phi^{\text{Dirac\ string}}+\Phi^{\text{line\ vortex}}$ represents the contribution of a line vortex equal to $+\dfrac{2\pi}{g}$ which enters the magnetic monopole sitting at the origin, $r=0$. 
As a result, the magnetic flux of the monopole is obtained as the sum of the absolute values of the fluxes of the two line vortices entering to it. Next, for the 
\textquotedblleft center projection\textquotedblright, we investigate the field strength tensor Eq.\eqref{12} by applying Eq.\eqref{23}. As a result, the first term  
$\vec{F}^{\text{linear}}_{\mu\nu}\equiv \partial_\mu \vec{A}^{\prime NM}_\mu-\partial_\nu\vec{A}^{\prime NM}_\mu$ contains some defects as the following
\begin{equation}\begin{split}
		&\left( \partial_\mu E_\nu-\partial_\nu E_\mu\right)T^3=\dfrac{1}{g}sin\theta\left(\partial_\mu\theta \partial_\nu\varphi-\partial_\mu\varphi\partial_\nu\theta \right)\ T^3 \\&+\dfrac{1}{g}\left( 1-cos\theta\right) \left[ \partial_\mu,\partial_\nu\right] \varphi\ T^3-\dfrac{1}{g} \left[ \partial_\mu,\partial_\nu\right] \varphi\ T^3.
		\label{24}
\end{split}\end{equation}
The first term of Eq.\eqref{24} represents the field strength of a magnetic monopole located at $r=0$, the second term indicates the field strength of a Dirac 
string at $\theta=\pi$ and the third term represents the field strength of a thin vortex field expanded on the $z$-axis. But the second term of 
Eq.\eqref{12}, $\vec{F}^{\text{bilinear}}_{\mu\nu}\equiv ig\left[ \vec{A}^{\prime NM}_\mu(x),\vec{A}^{\prime NM}_\nu(x)\right] $ is zero. Using the center gauge transformation defined in Eq.\eqref{7} and the Abelian gauge transformation of Eq.\eqref{17}, the third term of Eq.\eqref{12} is,
\begin{equation}
	\vec{F}^{\text{singular}}_{\mu\nu}.\vec{T}=-\dfrac{1}{g}\left( 1-cos\theta\right) \left[ \partial_\mu,\partial_\nu\right] \varphi\ T^3+\dfrac{1}{g} \left[ \partial_\mu,\partial_\nu\right] \varphi\ T^3,
	\label{25}
\end{equation}
where $-\dfrac{1}{g}\left( 1-cos\theta\right) \left[ \partial_\mu,\partial_\nu\right] \varphi\ T^3$ represents an anti-Dirac string in $\theta=\pi$ with a flux 
equal to $-\dfrac{4\pi}{g}T^3$ and the term $\dfrac{1}{g} \left[ \partial_\mu,\partial_\nu\right] \varphi\ T^3$ represents an anti-vortex on the $z$-axis 
with a flux equal to $+\dfrac{2\pi}{g}T^3$.
In fact, the contribution of the vortex and the Dirac string appearing in $\vec{F}^{\text{linear}}_{\mu\nu}$ is exactly canceled by the contribution of the anti-vortex and the anti-Dirac string in $\vec{F}^{\text{singular}}_{\mu\nu}$. As a result, a monopole vacuum is obtained unless we remove some of the singularities.
As explained in section \eqref{sec:level2}, a \textquotedblleft center projection\textquotedblright\ is done by removing $\vec{F}^{\text{singular}}_{\mu\nu}$ defined in Eq.\eqref{15}. This means that the 
center projection is obtained by $\vec{F}^{\text{linear}}_{\mu\nu}+\vec{F}^{\text{bilinear}}_{\mu\nu}$. 
On the other hand, we have shown that $\vec{F}^{\text{bilinear}}_{\mu\nu}$ is zero, and thus the center projected field strength tensor is equal to
$\vec{F}^{\text{CP}}_{\mu\nu}.\vec{T}=\left( \vec{F}^{\text{linear}}_{\mu\nu}.\vec{T}\right)$. Therefore, only a monopole attached to a Dirac string and a vortex are remained. We can interpret these configuration as a chain.

We end this section by discussing about the possible advantages of using chains.
None of the two models, the dual superconductor and the center vortex model, and their associated defects are able to describe all the expected features of the 
confining potential between color sources. We recall that the dependence of the potential slope to the Casimir scaling applies only for the intermediate distances and it is valid and exact for the large $N$ limit \cite{GH}. In addition, at large distances, the $k$-string tension depends on the $N$-ality of the representations. Vortex based models have been able to explain the $N$-ality dependence. However, to get the Casimir scaling for all representations, the models have been modified by defining a thickness to the vortex. On the other hand, lattice results confirm the existence of chains of monopoles and vortices that may explain the agreement of the potentials with Casimir scaling for higher representations.
In this article, motivated by lattice gauge theory results, we have introduced chains of monopoles and vortices for the continuum limit.

\section{\label{sec:level4}Conclusions}
In the direct method, by applying a center gauge transformation, we have shown that QCD vacumm is filled by vortices and anti-vortices. 
On the other hand, by the indirect method and applying two singular gauge fixing, the vortex and anti-vortex appear in the gauge theory along with the monopole. 
In fact, using the indirect method, we do not have single vortices but a chain that includes monopoles and vortices. Our results are in agreement with Del
Debbio's \textit{et al}.\cite{DFG97}, who have done lattice QCD calculations, as well as the results by
Engelhardt and Reinhardt \cite{ER2017}.



	
	
	



\end{document}